# Magnetic properties of Transition Metal Dimers probed by Inelastic Neutron Scattering

*Simon Ansbro[a,b], Eufemio Moreno-Pineda[c*], Wen Yu[c,d], Jacques Ollivier[b], Hannu Mutka[b], Mario Ruben[c,e] and Alessandro Chiesa[f,g] ** 

a. School of Chemistry, The University of Manchester, Oxford Road, Manchester, M13 9PL, UK.
b. Institut Laue-Langevin, 71 Avenue des Martyrs, Grenoble CS 20156, France.
c. Institute of Nanotechnology (INT), Karlsruhe Institute of technology (KIT), Hermann-von-Helmholtz-Platz 1, D-76344 Eggenstein-Leopoldshafen, Germany.
d. Hangzhou Wahaha Precision Machinery Co., Ltd, Street No. 14, Xiasha ETDZ, 310018 Hangzhou, China.
e. Institut de Physique et Chimie des Matériaux de Strasbourg (IPCMS), CNRS-Université de Strasbourg, 23 rue du Loess, BP 43, F-67034 Strasbourg Cedex 2, France.
f. Dipartimento di Scienze Matematiche, Fisiche e Informatiche, University of Parma, 43124 Parma, Italy.
g. Institute for Advanced Simulation, Forschungszentrum Jülich, Germany.

**ABSTRACT:** The physical characterisation and understanding of molecular magnetic materials is one of the most important steps towards the integration of such systems in hybrid spintronic devices. Amongst the many characterisation techniques employed in such a task, Inelastic Neutron Scattering (INS) stands as one of the most powerful and sensitive tools to investigate their spin dynamics. Herein, magnetic properties and spin dynamics of two dinuclear complexes, namely [(M(hfacac)$_2$)$_2$(bpym)] (where M = Ni$^{2+}$, Co$^{2+}$, abbreviated in the following as **Ni$_2$**, **Co$_2$**) are reported. These are model systems that could constitute fundamental units of future spintronic devices. By exploiting the highly sensitive IN5 Cold INS spectrometer, we are able to gain a deep insight into the spin dynamics of one **Ni$_2$** and to fully obtain the microscopic spin Hamiltonian parameters; while for **Co$_2$**, a multitude of INS transitions are observed demonstrating the complexity of the magnetic properties of octahedral cobalt-based systems.

*On occasion of the 60$^{th}$ anniversary of Prof. Kim Dunbar.*

**INTRODUCTION**

Molecular nanomagnets[1] (MNMs) are finite clusters of metal ions, whose spins are coupled by exchange interactions and are magnetically isolated from neighbouring ions by the surrounding bulky ligands. Thanks to their relatively small size and their zero-dimensional characteristics, they offer a remarkable opportunity to study fundamental physical phenomena of both quantum and classical nature.[1-8] Moreover, the high degree of achieved chemical control makes them promising systems for many different applications, such as quantum-information processing[9-14], high-density data storage[15-17] and low-temperature magnetic refrigeration[18,19].

In the last few years, important steps have been made in the field of molecular electronics, where the challenge to construct certain electrical or spintronic devices on a molecular scale is addressed by the assembly of molecules tailored for specific applications. In particular, molecule-scaled diodes, transistors and memories were realised[20-26] and the magnetoresistance effect was observed on single molecules[27,28]. In this respect, di-nuclear magnetic complexes are the optimal test-beds to understand the transport behaviour of MNMs. Indeed, magnetic intra-molecular as well as molecular-substrate interactions can be finely tuned by chemical engineering, thus allowing us to explore new features arising from the competition between these couplings.[29,30] Recently, a molecular nanomagnet dimer of chemical formula [(Ni(hfacac)$_2$)$_2$(bpym)] was deposited onto a copper surface, where it was demonstrated to display the Kondo effect for certain adsorption types, with a relatively high Kondo temperature of 10 K.[31] Electronic transport through a weakly coupled spin pair of $Co^{2+}$ ions was also reported, evidencing the switching of a Kondo-like anomaly by applying a bias voltage[30]. In turn, the characterisation of the magnetic properties of these complexes is essential to gain a deeper insight in their dynamics, thus paving the way for the rational design of future molecular devices. Indeed, the understanding of the correlation between structure and molecular properties is of the utmost importance, as to establish the design criteria for better performing molecular systems and their optimal integration in spintronic devices.

In this context, several characterisation techniques have been used for the understanding of the magnetic and spectroscopic properties of MNMs[1], ranging from conventional SQUID, μ-SQUID arrays, to spectroscopic techniques such Nuclear Magnetic Resonance (NMR), Electron Paramagnetic Resonance (EPR) and Inelastic Neutron Scattering (INS). INS is one of the best techniques available to characterise the microscopic magnetic interactions of MNMs[32-37]. It allows the precise determination of the parameters that define a spin Hamiltonian[36-37]. In addition, rapid developments in spectrometer technology make it possible to measure increasingly smaller samples, which is particularly useful where the chemical synthesis yields low quantities of product. Herein, we investigate the magnetic properties and the INS spectroscopic characteristics of two MNMs with formula [(M(hfacac)$_2$)$_2$(bpym)] (where M = Ni$^{2+}$, Co$^{2+}$), denoted as **Ni$_2$** and **Co$_2$**, for the Ni- and Co-containing molecules, respectively. The highly sensitive cold neutron spectrometer IN5 at the Institut Laue-Langevin allow us to directly probe the low-energy spin dynamics of the complexes on a very small amount of sample (ca. 0.1 g).

**EXPERIMENTAL SECTION**

*Synthesis and Crystallographic Analysis*

Complexes **Ni$_2$** and **Co$_2$** were obtained as described in published procedures.[31,38] Single crystals of **Ni$_2$** and **Co$_2$** were obtained from recrystallisation from a diethyl ether solution. Single crystal X-ray diffraction data of **Ni$_2$** and **Co$_2$** were collected employing a STOE StadiVari 25 diffractometer with a Pilatus300 K detector using GeniX 3D HF micro focus with MoKα radiation (λ = 0.71073 Å). The structure was solved using direct methods and was refined by full-matrix least-squares methods on all $F^2$ using SHELX-2014[39] implemented in Olex2[40]. The crystals were mounted on a glass tip using crystallographic oil and placed in a cryostream. Data were collected using ϕ and ω scans chosen to give a complete asymmetric unit. All non-hydrogen atoms were refined anisotropically. Hydrogen atoms were calculated geometrically riding on their parent atoms. The structures are isostructural with those earlier reported.[31,38] Full crystallographic details can be found in CIF format: see the Cambridge Crystallographic Data Centre database (CCDC 1023858 and 1143445).

*Magnetic measurements*

**Susceptibility and Magnetisation**. The magnetic susceptibility of compounds **Ni$_2$** and **Co$_2$** was measured in the temperature range 1.8 K – 300K employing powdered samples constrained in eicosane. The measurements were conducted using Quantum Design MPMS-XL SQUID magnetometer on polycrystalline material. Further magnetisation measurements were collected between 0 and 7 T in the temperature range of 1 and 10 K on a Quantum Design SQUID magnetometer MPMS®3 magnetometer equipped with a magnet operating between 0 and 7 T. The data were corrected for the diamagnetism of the compound (Pascal constants) and for the diamagnetic contribution of eicosane and the sample holder.

*INS Measurements*

Inelastic Neutron Scattering data for both compounds were collected on IN5 cold spectrometer at the Institute Laue-Langevin, with incident neutron wavelength 4.8 Å, corresponding to a resolution of 90 μeV in the temperature range 1.5-20 K.[41]

**RESULT AND DISCUSSION**

*Synthesis and Crystal structures*

The **Ni$_2$** and **Co$_2$** compounds were obtained by reacting one equivalent of bipyrimidine ligand with two equivalents of the M(hfacac)$_2$ (M = Ni$^{2+}$, Co$^{2+}$; hfacac = hexafluoroacetylacetonate) in a mixture of ethanol water.[31,38] Single crystal studies reveal the complexes to be two neutral dinuclear systems with formula [(Me(hfacac)$_2$)$_2$ (bpym)] (Fig. 1). Both complexes crystallise in the monoclinic *P2$_1$/c* space group, with the entire molecule in the asymmetric unit. In the structure, each transition metal centre, with an $N_2O_4$ coordination sphere, possesses a slightly distorted octahedral environment (with CShM values of 0.247 and 0.326 for both $Ni^{2+}$ ions for **Ni₂** and CShM values of 0.402 and 0.508 for both $Co^{2+}$ ions for **Co₂**, respectively).[42]

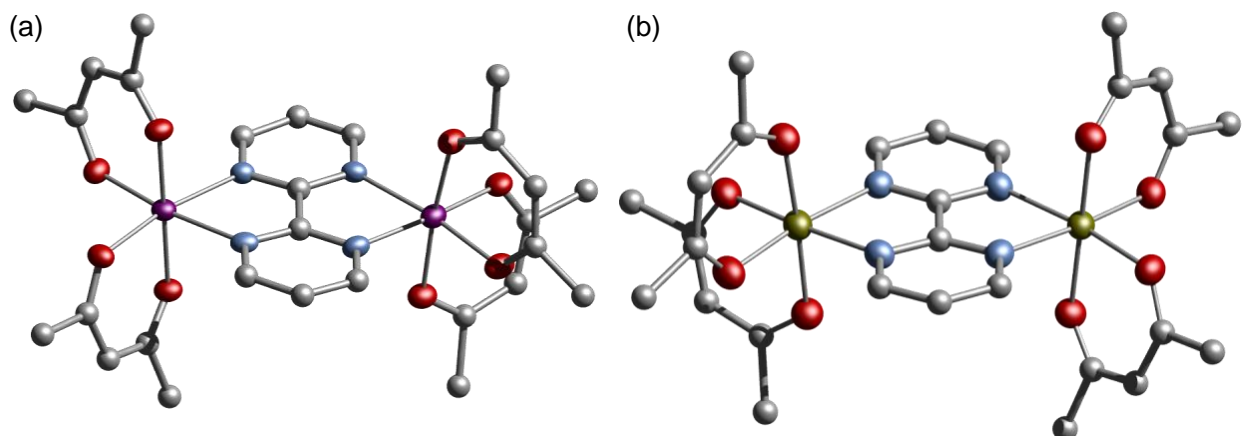

**Figure 1.** Crystals structure of **Co₂** (a) and **Ni₂** (b). Colour code: C, grey; N, cyan; O, red; Co, purple, Ni, green. Hydrogens and fluoride atoms were omitted for clarity.

The octahedral sites show very similar M-N and M-O distances: For **Ni₂**, the Ni⋯N distances vary from 2.087(3) to 2.113(3) Å, whilst for **Co₂** the Co⋯N distances range from 2.134(3) to 2.162(3) Å and the Co⋯O distances are between 2.036(2) to 2.059(2) Å. Due to the slightly shorter Ni⋯N distances, a shorter M⋯M distance is found for **Ni₂** with Ni⋯Ni of 5.603(3) Å, compared to **Co₂** where a Co⋯Co distance of 5.7268(8) Å is observed.

*Magnetic Description of Ni₂*

We begin our investigation of **Ni₂** from the magnetic properties, firstly exploring the temperature dependent static magnetic susceptibility behaviour $\chi_M(T)$ (where $\chi_M$ is the molar magnetic susceptibility), by measuring powdered samples under an applied field of 0.1 T. The compound shows a room temperature $\chi_M T(T)$ of 2.04 cm³ mol⁻¹ K (see Figure 2a). This value is consistent with two uncoupled $S = 1$ ions. $\chi_M T(T)$ stays practically constant with decreasing temperatures. This profile is observed up to ca. 40 K when it rapidly drops to zero, revealing a singlet ground state. Likewise, the $\chi_M(T)$ shows a maximum at approximately 24 K, indicative of the antiferromagnetic exchange between the two $Ni^{2+}$ centres or strong single ion anisotropy. The field dependent study, i.e. $M(H)$, from 2 to 5 K in the field range of 0 to 7 T, exhibits exponential-like profiles (Fig. 2b), characteristic of an $S = 0$ ground state with nearby excited states. The low temperature data clearly show that solely the singlet ground state is populated at 2 K. At higher temperatures and larger fields excited states are slowly populated, leading to an increased $M(H)$ profile. The magnetic properties, i.e. $\chi_M T(T)$ and $M(H)$, can be simultaneously fitted[43], employing a Hamiltonian of the following form:

$$\widehat{H} = -J\mathbf{s}_{Ni1} \cdot \mathbf{s}_{Ni1} + g_{Ni}\mu_B \mathbf{H} \cdot (\mathbf{s}_{Ni1} + \mathbf{s}_{Ni2}) \quad (1)$$

where the first term accounts for the isotropic exchange interaction between the $Ni^{2+}$ centres, and the last term is the Zeeman energy. Magnetisation and susceptibility are obtained as a function of temperature and field starting from the eigenvalues and eigenstates of the spin Hamiltonian. The best fits yield $J = -16$ cm$^{-1}$, with $g = 2.10$. Zero field splitting anisotropic terms have been omitted in (1), since we found that the $\chi_M T(T)$ and $M(H)$ are insensitive to these parameters. The here determined value of $J$ is in line with previous studies on exchange interactions between two $Ni^{2+}$ ions through a bpym bridging ligand. In particular, we find in our analysis a stronger exchange interaction if compared to Ref. 38 and very close to that reported in Ref. 44.

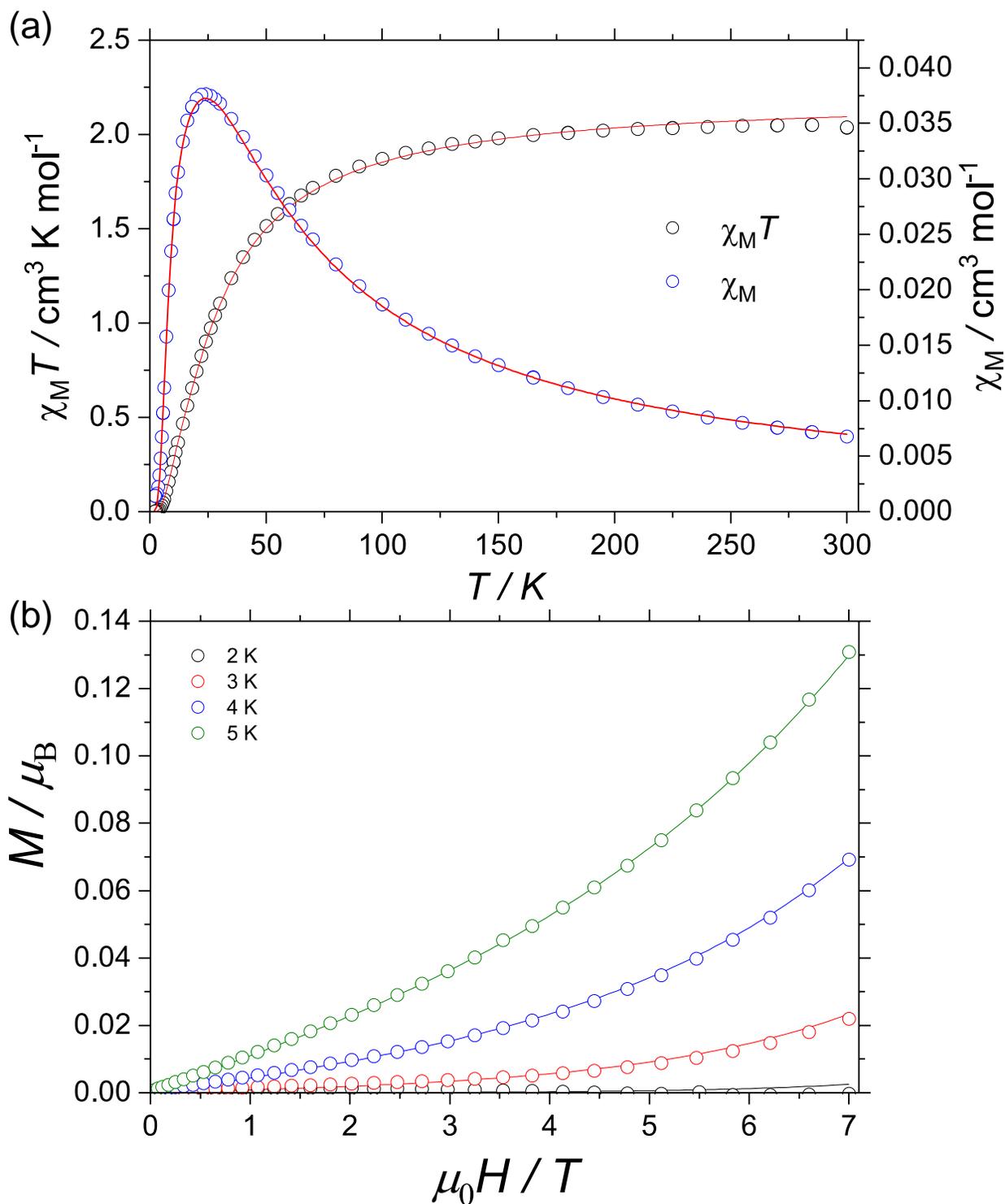

**Figure 2.** (a) Temperature dependent magnetic susceptibility data, as $\chi_M T(T)$ (black circles) and $\chi_M(T)$ (blue symbols) for **Ni₂**; (b) Experimental magnetisation vs. field for **Ni₂** at different temperatures. Solid lines are best fits employing Hamiltonian (1) and with the parameters described in the text. Simulation results obtained by employing Hamiltonian (1) and (2) are indistinguishable, evidencing that magnetometry in this case is not sensitive to anisotropy.

To further explore the magnetic properties of **Ni₂**, we turn our attention to the INS study carried out by exploiting the highly sensitive IN5 Cold spectrometer. Here we show that INS allows us not just to confirm the leading (exchange) interactions operating within the dimer, deduced from magnetic data, but also to probe smaller anisotropic terms that in the present case are not readily accessible from magnetometry. The INS spectra were collected on the small amount of 0.1 g of polycrystalline powders, by exploiting recent developments in spectrometer technology. Both energy and transferred momentum dependence of the scattered neutron intensity were analysed. This last information, obtained by integrating over the energy range corresponding to each magnetic transition, allows us to probe the structure of the involved eigenstates.

Figure 3 shows INS spectra vs. transferred energy for **Ni₂** measured at three different temperatures, 1.6, 10 and 20 K. At low temperature, three peaks are clearly distinguishable at 14.8 cm$^{-1}$ (I), 16.5 cm$^{-1}$ (II) and 18.2 cm$^{-1}$ (III) (Fig 3). These three transitions correspond to cold inter-multiplet transitions between the ground state singlet and the first excited triplet (split by anisotropy). Two other broad transitions are also observed at low energy, i.e. 1.7 cm$^{-1}$ (i) and 3.3 cm$^{-1}$ (ii), whose intensity increases upon increasing temperatures.

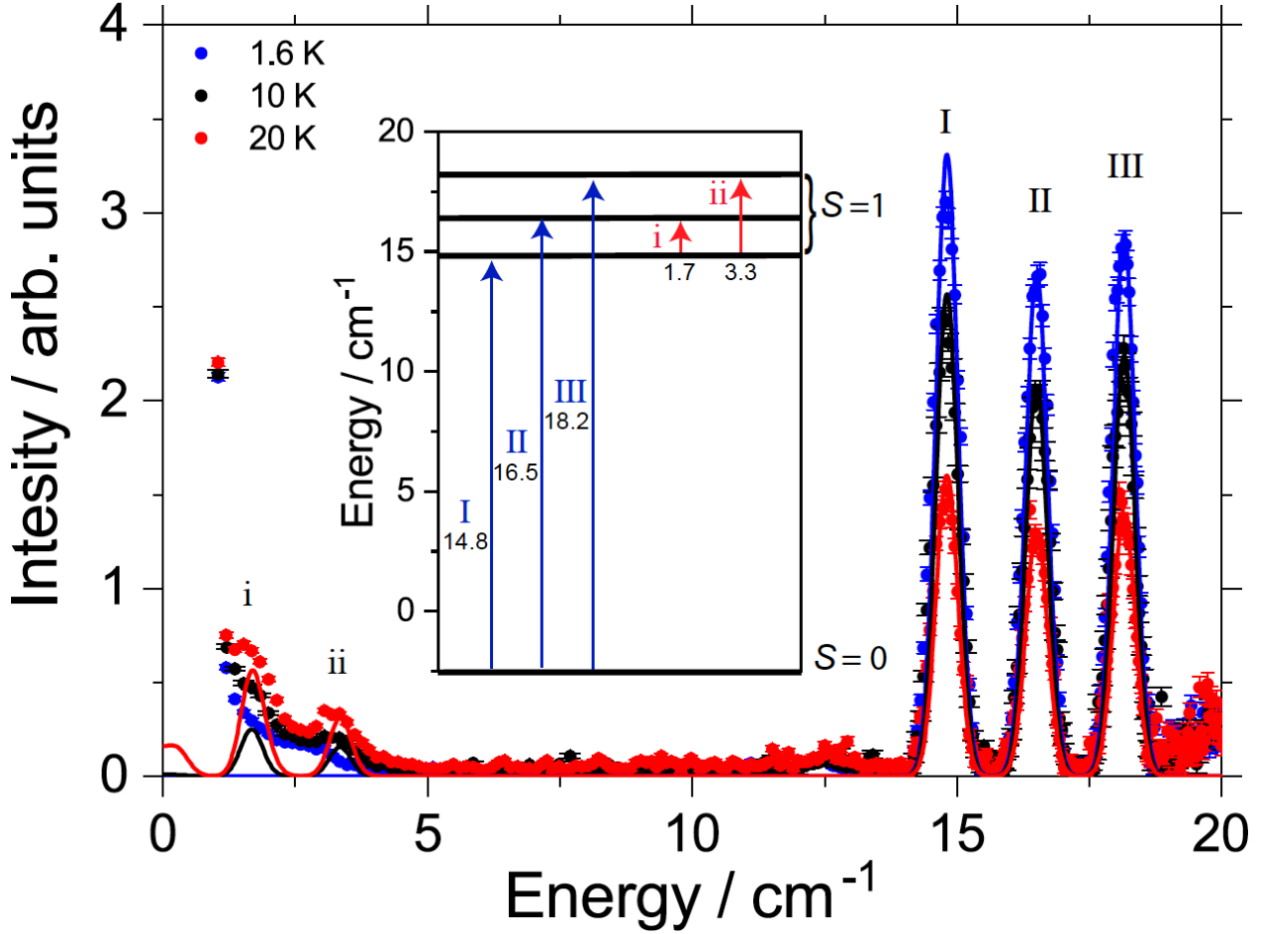

**Figure 3.** Measured (symbols) energy spectra for **Ni₂** at 1.6, 10 and 20 K, and corresponding simulations (lines) calculated by using Eq. (5), with eigenvalues and eigenstates obtained from diagonalization of Hamiltonian (2). The inset reports the level diagram for the states involved in the cold (blue arrows) and hot (red) transitions probed by INS, with indicated excitation energies and approximate total spin $S$.

To account for the structure of the observed excitations, the INS data have been fitted using the microscopic spin Hamiltonian of the form (2), where the axial and rhombic zero field splitting terms are introduced:

$$\hat{H} = -J\mathbf{s}_1 \cdot \mathbf{s}_2 + d\sum_{i=1}^{2}[s_{zi}^2 - s(s+1)/3] + e\sum_{i=1}^{2}(s_{xi}^2 - s_{yi}^2) \qquad (2)$$

Here $i = 1, 2$ labels the two magnetic ions, $J$ is the isotropic exchange interaction, while $d$ and $e$ are the axial and rhombic zero-field splitting anisotropy parameters, respectively. Since the two ions are essentially related by an inversion centre, we have modelled them with the same zero-field splitting parameters. The Zeeman term is omitted because measurements were carried out at zero field. In order to reduce the number of free parameters, we have not included anisotropic exchange terms, which are usually negligible compared to zero-field splitting for Ni$^{2+}$ complexes in an octahedral environment[45].

It is useful to obtain analytical expressions of the molecular eigenstates belonging to different total spin multiplets (eigenstates of the isotropic, leading term of the spin Hamiltonian), and then consider the effect of anisotropic terms as a perturbation. This corresponds to finding the transformation from the product ($|m_1, m_2\rangle$, with $m_i$ the eigenvalues of $s_{zi}$) to the total spin basis ($|\psi_{S,M}\rangle$, being $S$ the total spin and $M$ its third component). We assume $z$ as the quantisation axis.

In particular, the ground state can be expressed as:

$$|\psi_{0,0}\rangle = \frac{1}{\sqrt{3}}(|1,-1\rangle + |-1,1\rangle - |0,0\rangle) \quad (3)$$

while for the first excited triplet (characterised by excitation energy $J$) one easily finds $|\psi_{1,0}\rangle = \frac{1}{\sqrt{2}}(|1,-1\rangle - |-1,1\rangle)$ and $|\psi_{1,\pm 1}\rangle = \frac{1}{\sqrt{2}}(|\pm 1,0\rangle - |0,\pm 1\rangle)$. $|\psi_{1,0}\rangle$ and $|\psi_{1,\pm 1}\rangle$ are split to first order by single-ion axial anisotropy terms. This giant $S = 1$ experiences an effective $D = -d$, thus a positive microscopic $d$ results in an effective easy axis anisotropy. The two degenerate $|\psi_{1,\pm 1}\rangle$ states are then further split by rhombic single ion terms into symmetric and anti-symmetric combinations:

$$|\psi_{1,\pm}\rangle = \frac{1}{\sqrt{2}}\left(|\psi_{1,+1}\rangle \pm |\psi_{1,-1}\rangle\right) = \frac{1}{2}(|1,0\rangle - |0,1\rangle \pm |-1,0\rangle \mp |0,-1\rangle) \quad (4)$$

with gap 2$e$. This first order perturbative analysis leads directly to the expressions for the position of the three main peaks, given by: $J - e - d/3$, $J + e - d/3$ and $J + 2d/3$. Since these are approximately equally spaced we obtain $e \approx d/3$. Notice that the sign of the anisotropy parameters does not alter the form of the eigenstates, but only changes their order (see discussion below).

The inelastic neutron cross-section was numerically simulated by using the formula[46]

$$S(\mathbf{Q},\omega) \propto \sum_{\lambda,\lambda'} \frac{e^{-\beta E_\lambda}}{Z} I_{\lambda,\lambda'}(\mathbf{Q}) \, e^{-\frac{\left(\hbar\omega + E_\lambda - E_{\lambda'}\right)^2}{2\sigma^2}} \quad (5)$$

where $|\lambda\rangle$ are molecular eigenstates with energy $E_\lambda$, **Q** and $\omega$ are the transferred momentum and energy, $\beta = 1/k_BT$, Z is the partition function and $\sigma$ is the standard deviation of the Gaussian line-shape associated with each transition (given by the instrumental resolution). Here

$$I_{\lambda,\lambda'}(\mathbf{Q}) = \sum_{ij} F_i(Q) F_j^*(Q) \, e^{i\mathbf{Q}\cdot\mathbf{R}_{ij}} \sum_{\alpha\beta} \left(\delta_{\alpha\beta} - \frac{Q_\alpha Q_\beta}{Q^2}\right) \langle\lambda|s_{\alpha i}|\lambda'\rangle \langle\lambda'|s_{\beta j}|\lambda\rangle. \quad (6)$$

The cross-section for a powder sample is obtained by averaging $I_{\lambda,\lambda'}(\mathbf{Q})$ over all the possible directions of **Q**. This average can be analytically determined as shown in Ref. 47:

$$I_{\lambda,\lambda'}(Q) = \sum_{i,j=1,2} F_i^*(Q) F_j(Q) \Big\{ \tfrac{2}{3}[j_0(QR_{ij}) + C_0^2 j_2(QR_{ij})]\langle\lambda|s_{zi}|\lambda'\rangle\langle\lambda'|s_{zj}|\lambda\rangle +$$
$$\tfrac{2}{3}\left[j_0(QR_{ij}) - \tfrac{1}{2}C_0^2 j_2(QR_{ij})\right](\langle\lambda|s_{xi}|\lambda'\rangle\langle\lambda'|s_{xj}|\lambda\rangle + \langle\lambda|s_{yi}|\lambda'\rangle\langle\lambda'|s_{yj}|\lambda\rangle) +$$
$$\tfrac{1}{2}j_2(QR_{ij})C_2^2(\langle\lambda|s_{xi}|\lambda'\rangle\langle\lambda'|s_{xj}|\lambda\rangle - \langle\lambda|s_{yi}|\lambda'\rangle\langle\lambda'|s_{yj}|\lambda\rangle) \Big\} \quad (7)$$

where $F_i(Q)$ is the magnetic form factor for ion $i$, $\mathbf{R}_{ij}$ gives the relative position of ions $i$ and $j$, $j_{0,2}(QR_{ij})$ are spherical Bessel functions. Here $C_0^2 = \tfrac{1}{2}\left[3\left(\frac{R_{ijz}}{R_{ij}}\right)^2 - 1\right]$, $C_2^2 = \frac{R_{ijx}^2 - R_{ijy}^2}{R_{ij}^2}$.

In Eq. (7) we have not included matrix elements $\langle\lambda|s_{\alpha i}|\lambda'\rangle\langle\lambda'|s_{\beta j}|\lambda\rangle$ with $\alpha \neq \beta$, which in the present case do not contribute to the cross-section.

Numerical simulations of the scattered cross-section [Eq. (5)] with eigenstates obtained from diagonalization of Hamiltonian (2) yield the refined parameters $J = -16.34$ cm$^{-1}$, $d = 2.53$ cm$^{-1}$ and a high $e = 0.84$ cm$^{-1}$. This set of parameters also reproduces the low-energy excitations, occurring approximately at energies $2e, d - e$ [almost degenerate, broad peak (i)] and $d + e$ (ii). As correctly described by our model, these are transitions from an excited state (whose population increases with temperature).

While the absolute values of the spin Hamiltonian parameters $J$, $d$ and $e$ can be precisely determined from the position of peaks in the $S(\omega)$ spectrum, information on the excited eigenstates can be deduced from the relative intensities of the cold excitations I, II and III in the energy spectrum and especially from the Q-dependent spectrum[48]. This, in turn, allows us to fix the direction of the anisotropy axes with respect to the Ni⋯Ni bond.

The extracted Q-dependence for the three main peaks is shown in Figure 4, with the typical pattern of maxima and minima arising from interference terms between the two sites ($i \neq j$) in Eq. (7). We note that transitions I and III display a similar behaviour, with the first maximum at about 0.75 Å$^{-1}$, while for transition II the first peak occurs at ca. 0.95 Å$^{-1}$. This behaviour, as long as the relative intensities of peaks I, II and III in the energy spectrum, can only be reproduced by assuming *d* and *e* with the same sign and the Ni⋯Ni bond along the intermediate anisotropy axis (i.e. *x* axes if *d, e* > 0 or, equivalently, *y* if *d, e* < 0). This can be verified by applying Eq. (7) to the approximate eigenstates $|\lambda\rangle = |\psi_{0,0}\rangle$ and $|\lambda'\rangle = |\psi_{1,0}\rangle, |\psi_{1,\pm}\rangle$. Conversely, the sign of *d* cannot be determined from the data. The good agreement between measured and simulated Q-dependence of the scattered intensity (obtained from full diagonalization of Hamiltonian (2)) reinforces the reliability of the fitted parameters and confirms the structure of the eigenstates.

The parameters obtained through INS data also finely reproduce the magnetometry data, which is insensitive to the rhombic and axial zero-field field parameters (*vide supra*). Note that inclusion of *d* in the Hamiltonian, based solely upon magnetic data, i.e. $\chi_M T(T)$, yields a large overestimation of *d* (31 cm$^{-1}$ in ref. 38 cf. 2.53 cm$^{-1}$ in this work).

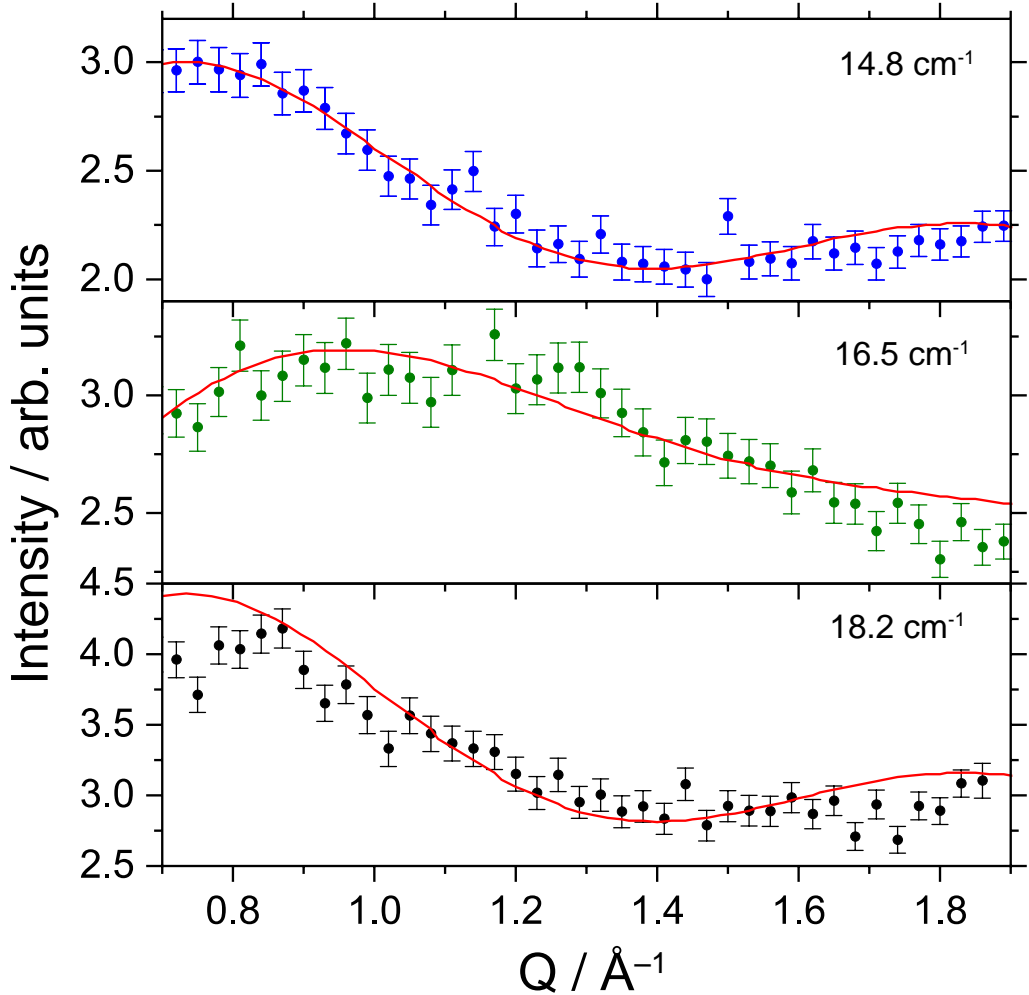

**Figure 4.** Measured (symbols) and simulated (curves) Q-dependence for peaks I, II and III at $T = 1.6$ K for **Ni$_2$**.

*Magnetic description of Co$_2$*

**Co$_2$** shows a room temperature value of 5.41 cm$^3$ mol$^{-1}$ K. Upon cooling the $\chi_M(T)T$ declines smoothly up to approximately 40 K, where it sharply drops to a $\chi_M T$ value of 0.37 cm$^3$ mol$^{-1}$ K at 2 K (Fig. 5a). This behaviour is consistent with the orbitally degenerate $^4T_{1g}$ ground term of hexa-coordinated Co$^{2+}$, with a well-isolated spin–orbit doublet ground state. Moreover, inspection of the $\chi_M(T)$ plot shows a maximum at ca. 6.8 K, which decreases upon lowering temperatures. This maximum can again be ascribed to antiferromagnetic exchange operating within the metal Co$^{2+}$ centres. The magnetisation shows an almost linear increase with the applied field, with a very small magnetisation value reaching 2.5 $\mu_B$ (inset of Fig. 5a).

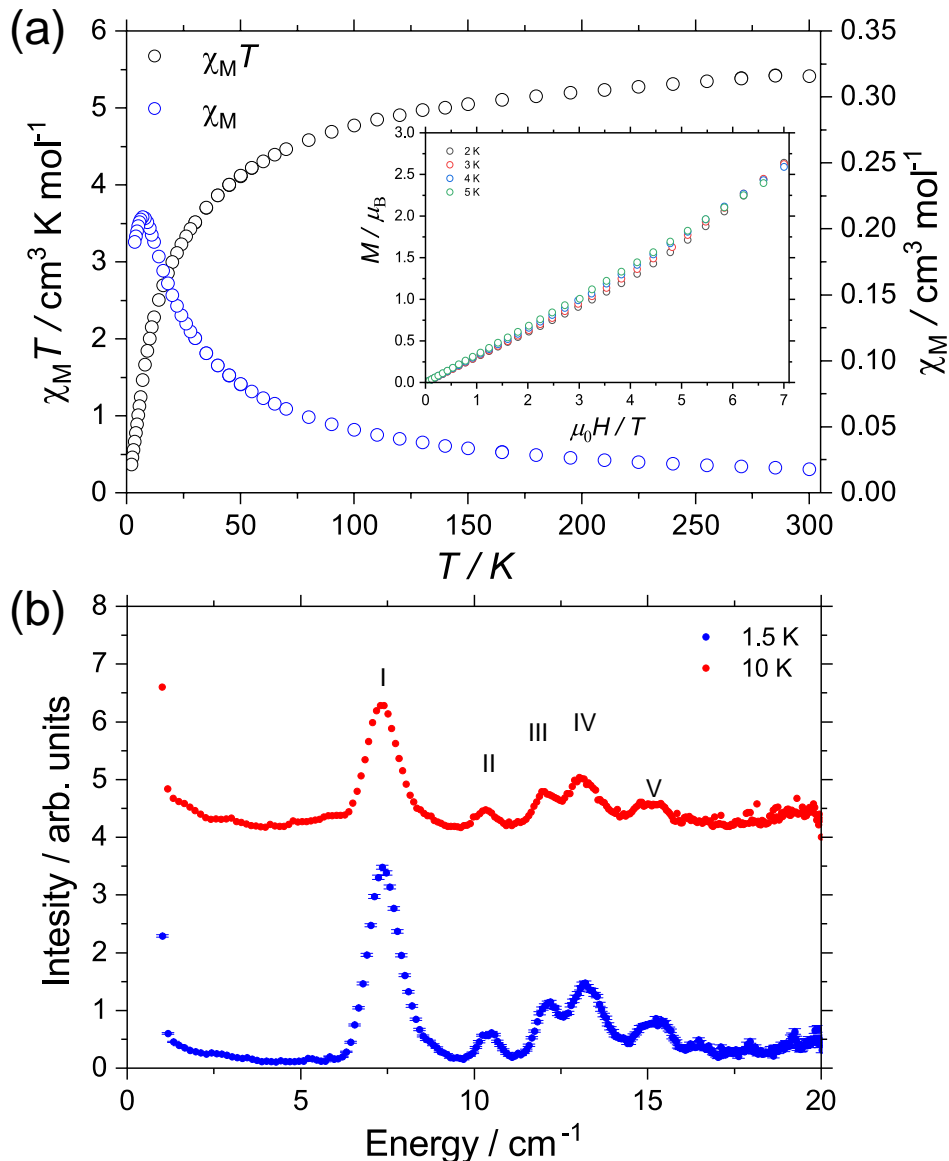

**Figure 5.** Temperature-dependence of the susceptibility, field-dependence of the magnetisation (a) and energy spectra at 1.5 and 10 K (b) for **Co₂**, collected with incident neutron wavelength of 4.8 Å. Data of INS spectra can be found in the SI.

INS spectra for **Co₂** at temperature of 1.5 and 10 K using a wavelength of 4.8 Å are shown in Fig. 5b. These are more complicated than that observed for **Ni₂** and much more informative than magnetometry data. In particular, we note that the presence of at least five peaks (some of them probably split) between 7.3 cm$^{-1}$ and 16.1 cm$^{-1}$ in the low-temperature INS spectrum can only be explained by assuming the Co$^{2+}$ characterised by an effective spin $^3/_2$. Indeed, since all the detected excitations occur at energies $\gg \kappa_B T$, some levels (two at least) must be not populated at 1.5 K. This reduces the possible number of peaks (in the case of two interacting doublets) to four. Hence, INS data demonstrate that Co$^{2+}$ is here in its high spin ($s = ^3/_2$) state. By comparing the two sets of measurements collected at 1.5 and 10 K (blue and red dots in Fig. 5b, respectively), we note that the intensity of all the peaks decreases by increasing temperature, as expected for cold transitions from the ground to excited states. However, no additional peaks are measured at 10 K at low energy, despite the significant population of the excited state at 7.3 cm$^{-1}$ ($\approx$10 K). This suggests a low transition probability between the level at 7.3 cm$^{-1}$ and those immediately higher in energy.

Despite an extensive exploration of the parameters space, it was not possible to find a univocal set of parameters to account for the observed behaviour of susceptibility, *M(H)* and INS data. Further studies (e.g. by using the recently developed four-dimensional INS technique[34-37]) would be needed to access the structure of the molecular eigenstates, thus obtaining a sound characterisation of the microscopic spin Hamiltonian.

**CONCLUSIONS**

We have used INS in combination with SQUID magnetometry to investigate two binuclear molecular nanomagnet complexes, namely **Ni₂** and **Co₂**, obtaining the full microscopic spin Hamiltonian for the former. The Q-dependent spectrum has been particularly useful to determine the orientation of the anisotropy axes. By exploiting recent advances in the spectrometer technology, we could perform measurements on small quantities of sample, which is very promising for the development of molecular electronic devices, where frequently the chemical yield of the synthesis is low. Systems such as this dimer provide excellent models for gaining crucial insights into quantum transport properties through molecular junctions. The detailed characterisation of these systems provided by INS, which gives access also to subtle anisotropic terms in the spin Hamiltonian, is fundamental for the development of future technologies, based on molecular transistor setups, which have been demonstrated to be highly selective with respect to spin anisotropy[49].

**AUTHOR INFORMATION**

**Corresponding Authors**

*E-mails: alessandro.chiesa@unipr.it; eufemio.pineda@kit.edu

**Supporting Information Available**: further synthetic details, structural and magnetic plots.

**Conflicts of interest**

The authors report no conflicts of interest to declare.

**ACKNOWLEDGEMENTS**

We acknowledge the support of the Deutsche Forschungsgemeinschaft (DFG, TRR 88, "3MET", project A8) and FET-Open project "MOQUAS" by the European Commission, and the French Government through the ANR project "MolQuSpin". AC acknowledges "Fondazione Angelo Della Riccia" for financial support and S. Carretta for useful discussions.

## For Table of Contents Only

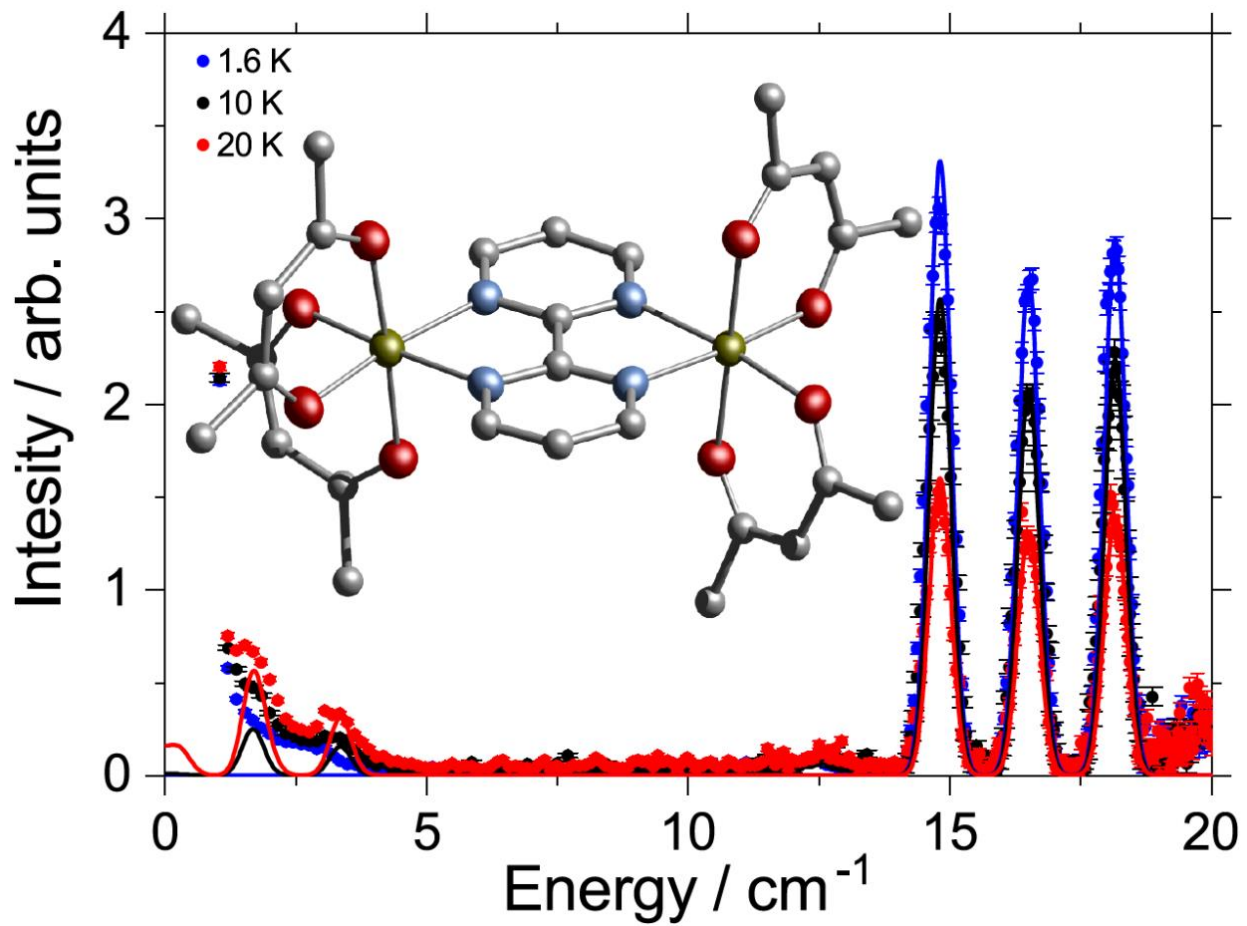

We characterise the spin dynamics of two molecular transition metal dimers, by combining Inelastic Neutron Scattering and magnetic measurements.